\documentclass{emulateapj}
\usepackage{graphicx}

\begin{document}

\title{High--Resolution Spectroscopic Imaging of CO in a z = 4.05 Proto--cluster}

\shorttitle{GN20.2a/b}

\author{J. A. Hodge\altaffilmark{1}}
\altaffiltext{1}{Max--Planck Institute for Astronomy, K\"{o}nigstuhl 17, 69117 Heidelberg, Germany}
\email{hodge@mpia.de}

\author{C. L. Carilli\altaffilmark{2,3}}
\altaffiltext{2}{National Radio Astronomy Observatory, P.O. Box 0, Socorro, NM 87801-0387, USA}
\altaffiltext{3}{Astrophysics Group, Cavendish Laboratory, JJ Thomson Avenue, Cambridge CB3 0HE, UK}

\author{F. Walter\altaffilmark{1,2}}

\author{E. Daddi\altaffilmark{4}}
\altaffiltext{4}{CEA, Laboratoire AIM--CNRS--Universit\'{e} Paris Diderot, Irfu/SAp, Orme des Merisiers, F-91191 Gif--sur--Yvette, France}

\author{D. Riechers\altaffilmark{5,6}}
\altaffiltext{5}{Department of Astronomy, California Institute of Technology, MC 249-17, 1200 East California Boulevard, Pasadena, CA 91125, USA}
\altaffiltext{6}{Cornell University, 220 Space Sciences Building, Ithaca, NY 14853, USA}

\begin{abstract}
\noindent We present a study of the formation of clustered, massive galaxies at large look--back times via spectroscopic imaging of CO in the unique GN20 proto--cluster at z $=$ 4.05.
Existing observations show that this is a dense concentration of gas--rich, very active star forming galaxies, including multiple bright submillimeter galaxies (SMGs).
Using deep, high--resolution VLA CO(2--1) observations,
we image the molecular gas with a resolution of $\sim$1 kpc just 1.6 Gyr after the Big Bang. 
The SMGs GN20.2a and GN20.2b have deconvolved sizes of $\sim$5 kpc$\times$3 kpc and $\sim$8 kpc$\times$5 kpc (Gaussian FWHM) in CO(2--1), respectively, 
and we measure gas surface densities up to $\sim$12,700/1,700$\times$(sin \textit{i}) ($\alpha_{\rm CO}$/0.8) M$_{\odot}$ pc$^{-2}$ for GN20.2a/GN20.2b in the highest--resolution maps.
Dynamical mass estimates allow us to constrain the CO--to--H$_2$ conversion factor
to $\alpha_{\rm CO} =$ 1.7$\pm0.8$ M$_{\odot}$ (K km s$^{-1}$ pc$^{2}$)$^{-1}$ for GN20.2a and 
$\alpha_{\rm CO} =$ 1.1$\pm^{1.5}_{1.1}$ M$_{\odot}$ (K km s$^{-1}$ pc$^{2}$)$^{-1}$ for GN20.2b.
We measure significant offsets (0.5$^{\prime\prime}$--1$^{\prime\prime}$) between the CO and optical emission, indicating either dust obscuration on scales of tens of kpc or that the emission originates from distinct galaxies.
CO spectral line energy distributions imply physical conditions comparable to other SMGs and reveal further evidence that GN20.2a and GN20.2b are in different merging stages.
We carry out a targeted search for CO emission from the 14 known $B$--band Lyman break galaxies in the field, tentatively detecting CO in a previously--undetected LBG
and placing 3$\sigma$ upper limits on the CO luminosities of those that may lie within our bandpass.
A blind search for emission line sources 
down to a 5$\sigma$ limiting CO luminosity of $L^{\prime}_{\rm CO(2-1)} = 8 \times 10^{9}$ K km s$^{-1}$ pc$^{2}$ and covering $\Delta z = 0.0273$ ($\sim$20 comoving Mpc)
produces no other strong contenders associated with the proto--cluster.


\end{abstract}

\section{INTRODUCTION}
\label{Intro}

Massive elliptical galaxies seen in the local universe are believed to have descended from a population of intensely star--forming galaxies at high ($z$$>$2) redshift.
The best candidates are submillimeter galaxies \citep[SMGs;][]{2002PhR...369..111B}, gas--rich galaxies whose 
many young stars heat the surrounding dust and
cause them to be extremely bright in the submillimeter regime.
SMGs are forming stars at exceptionally high rates \citep[$\sim$10$^{3}M_{\odot}$ yr$^{-1}$; e.g.,][]{2003AJ....125..383A, 2005ApJ...632..736A, 2009ApJ...699.1610H}, resulting in huge bolometric luminosities of $\sim$10$^{13}$ $L_{\odot}$. 
Their redshift distribution is thought to peak around $z \sim 2.3-2.4$ \citep[e.g.,][]{2003Natur.422..695C},
though there is mounting evidence for a high--redshift tail extending above $z > 4$ \citep[e.g.,][]{2008ApJ...681L..53C,2008ApJ...689L...5S,2009ApJ...694.1517D, 2010ApJ...720L.131R,2011Natur.470..233C, 2011MNRAS.415.1479W,2012Natur.486..233W,2013ApJ...767...88W, 2013Natur.496..329R}.

GN20, so named after the GOODS--N field where it was originally discovered \citep{2006MNRAS.370.1185P}, is one such high--z SMG. 
A serendipitous detection of its CO(4--3) emission with the PdBI \citep{2009ApJ...694.1517D} established its redshift as $z = 4.05$, only 1.6 Gyr after the Big Bang. 
With a 350 GHz flux density of 20.3 mJy, it is 
the brightest SMG in the GOODS--N field and
the most luminous starburst galaxy known at $ z > 4$ \citep{2006MNRAS.370.1185P}.
A detailed study of the gas dynamics and morphology of GN20 was presented in \citet{2012ApJ...760...11H},
showing at high--resolution that GN20 contains an extended, clumpy, rotating gas disk. 

\begin{figure*}
\centering
\includegraphics[scale=0.95]{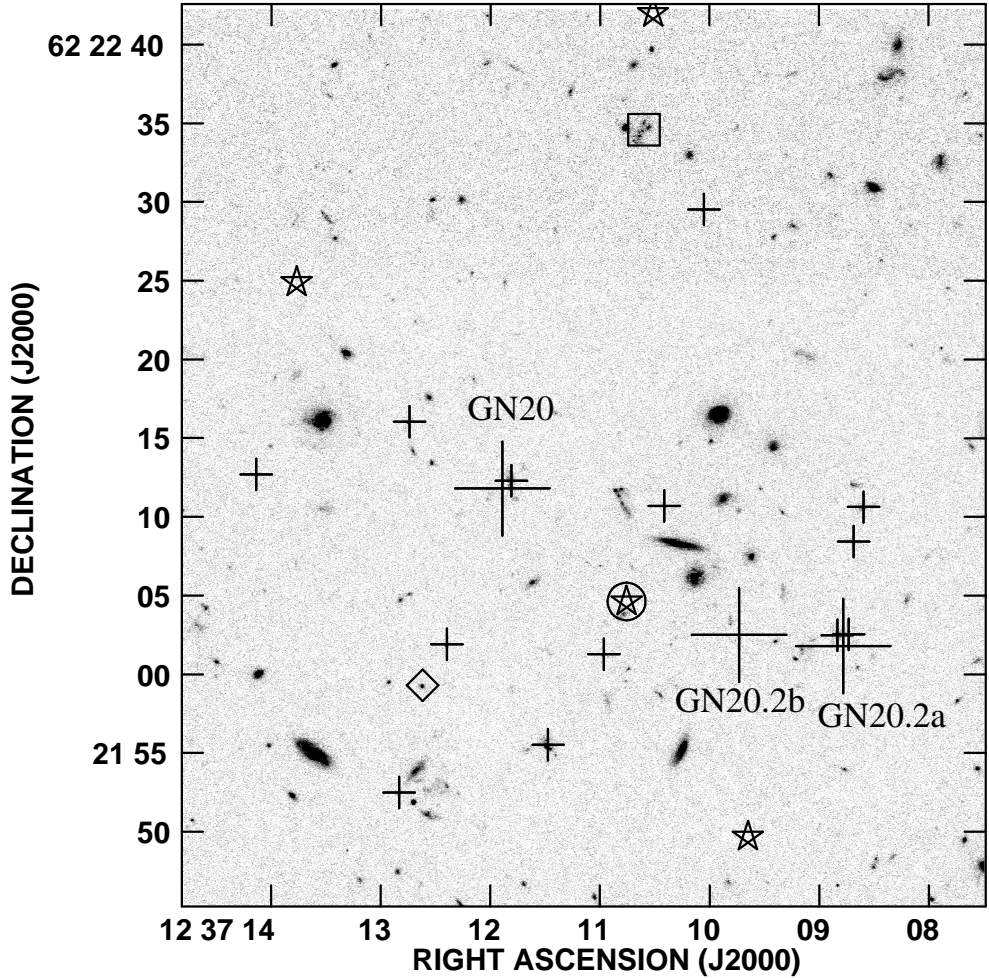}
\caption{HST$+$ACS z850--band image of the GN20 proto--cluster field. Large crosses mark the 1.4 GHz positions of the three known SMGs in the field. Small crosses mark the positions of the LBGs within 25$^{\prime\prime}$ of the SMG GN20, and the diamond marks the LBG with a possible detection in CO(2--1) (this work). The stars mark the positions of the new CO emission line source candidates (this work), and the circled star is the emission line candidate with an optical counterpart within 1$^{\prime\prime}$. The box shows the position of the $z=1.5$ galaxy BzK-21000. }
\label{fig:field}
\end{figure*}

What makes GN20 even more special is that it seems to lie in a dense concentration of galaxies (Figure~\ref{fig:field}).
Two additional SMGs, originally detected as a single source in the low--resolution SCUBA image \citep[`GN20.2'][]{2006MNRAS.370.1185P}, lie within $\sim$25$^{\prime\prime}$ of GN20.  
This corresponds to a projected physical separation of only $\sim$170 kpc.
These SMGs, referred to as GN20.2a and GN20.2b, have a separation of only a few arcseconds and redshifts of $z =$ 4.059 and 4.052, respectively \citep{2009ApJ...694.1517D, 2011ApJ...739L..33C}, very close to the redshift of GN20.
Their combined IR luminosity is 1.6 $\times$ 10$^{13}$ L$_{\odot}$ \citep{2009ApJ...694.1517D}, the equivalent of a hyper--luminous infrared galaxy with a SFR $>$1000 M$_{\odot}$ yr$^{-1}$.
A fourth SMG, GN10, is only a few arcminutes away, and with a CO--derived redshift of $z = 4.0424$, may also be related \citep{2009ApJ...695L.176D}.

Along with the multiple $z\sim4.05$  SMGs, the field also contains numerous Lyman Break Galaxies ($B$-band dropouts at $z \sim 4$), 14 of which lie within 25$^{\prime\prime}$ (2.5 comoving Mpc, projected) of GN20. 
This corresponds to an overdensity of 5.8$\sigma$ \citep{2009ApJ...694.1517D}.
The overdensity of $z > 3.5$ IRAC selected galaxies is a factor of 18, corresponding to a chance probability of 10$^{-4}$ assuming no spatial clustering.
The data therefore supports a very strong overdensity at $z \sim$ 4 with a transverse size of 2 comoving Mpc ($\Delta z \sim 0.0028$) and a total mass of $\sim$10$^{14}$ M$_{\odot}$, suggesting the presence of a massive proto--cluster (Figure~\ref{fig:field}).
Previous work has speculated that such overdense environments may play a role in triggering extreme star formation (\citealt{2007A&A...468...33E,2009ApJ...694.1517D}; but c.f.\ \citealt{2009ApJ...691..560C}).
The GN20 field is therefore a prime target for the study of galaxy and cluster formation in the early universe. 

One of the key observables in this case is the material which fuels the star formation: i.e., the molecular gas.  
As molecules of hydrogen (the most abundant species of molecular gas) have no permanent dipole moment, this is done through observations of the rotational transitions of carbon monoxide (CO). 
Observations of the CO emission, along with a carefully--chosen CO luminosity--to--H$_{2}$ gas mass conversion factor, provide crucial information on the amount of material available for star formation and its dynamical state \citep[e.g.,][]{2013arXiv1301.0371C}.	
The brightness temperature ratios from the different rotational transitions of CO can shed light on the temperature and density of the gas, as well as on the heating and excitation source.  
Lower--J transitions are particularly important, as they are thought to trace the cold molecular gas making up the bulk of the systems.
This enables more robust estimates of both the gas mass as well as the dynamical mass, as the full size of the reservoir is more reliably traced.

\begin{figure*}
\centering
\includegraphics[scale=0.65]{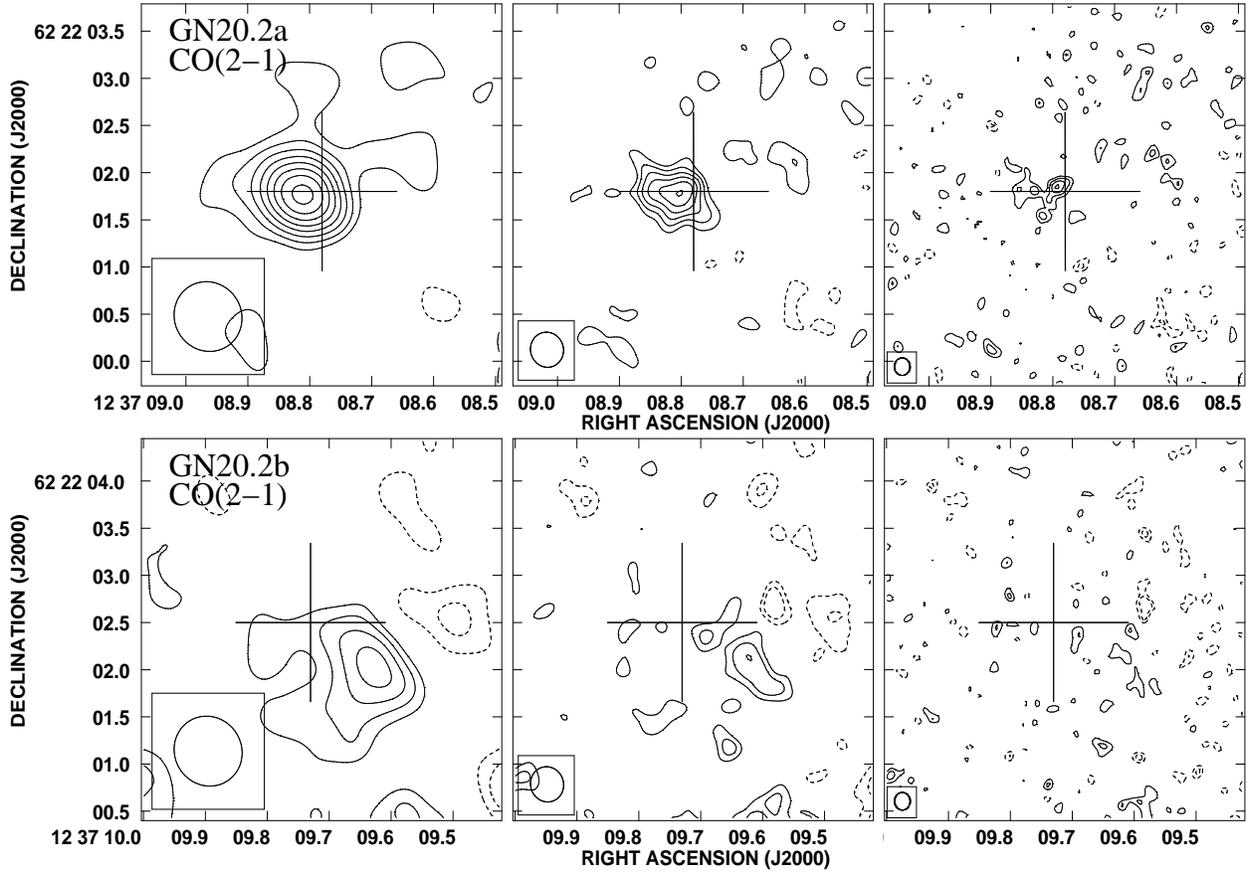}
\caption{Velocity--averaged CO(2--1) images of GN20.2a (top, over 780 km s$^{-1}$) and GN20.2b (bottom, over 470 km s$^{-1}$).  The maps have angular resolutions of (from left to right) 0.77$^{\prime\prime}$ (5.4 kpc), 0.38$^{\prime\prime}$ (2.7 kpc), and 0.19$^{\prime\prime}$ (1.3 kpc). The crosses show the position of the 1.4 GHz counterparts at 1.7$^{\prime\prime}$ resolution \citep[as indicated by the extent of the cross;][]{2010ApJS..188..178M}.  The maps have been primary beam corrected, and the rms noise values for GN20.2a are 26.0 $\mu$Jy beam$^{-1}$, 20.0 $\mu$Jy beam$^{-1}$, and 17.5 $\mu$Jy beam$^{-1}$, respectively. The rms noise values for GN20.2b are 28.0 $\mu$Jy beam$^{-1}$, 23.3 $\mu$Jy beam$^{-1}$, and 21.2 $\mu$Jy beam$^{-1}$. For all maps, the contours are shown in steps of 1$\sigma$ starting at $\pm$2$\sigma$.}
\label{fig:GN20_2_maps}
\end{figure*}

In \citet{2012ApJ...760...11H}, we presented a detailed analysis of the CO(2--1) emission in the SMG GN20 using a uniquely deep, high--resolution dataset.  
Here, we have used that same dataset (as well as some ancillary data) to study the molecular gas emission from the other sources in the GN20 proto--cluster field.
We begin in Section~\ref{obs} by introducing the different observational data used in this paper.
In Sections~\ref{gn20_2a_results} and \ref{gn20_2b_results}, we present the results of the observations for the SMGs GN20.2a and GN20.2b.
Section \ref{lbgs} describes the results of a targeted search for CO emission from proto--cluster LBGs,
and Section \ref{serch} describes a blind search for CO emission--line sources.
We present our analysis for GN20.2a and GN20.2b in Section~\ref{GN20_2_anal}, 
including CO sizes and gas surface densities (Section~\ref{COsize}),
star formation rate surface densities 
(Section~\ref{sfe}),
dynamical mass estimates and constraints on the CO--to--H$_{2}$ conversion factor (Section~\ref{Mdyn}),
localization of the counterparts (Section~\ref{localization}),
and spectral line energy distributions (Section~\ref{sled}).
We then briefly discuss 
the implications of the blind and targeted searches for CO emission from other sources in the field (Section~\ref{field}).
We end with our conclusions in Section~\ref{conclusions}.
Where applicable we assume the standard $\Lambda$ cosmology of H$_0$ = 70 km s$^{-1}$ Mpc$^{-1}$, $\Omega_{\Lambda}$ = 0.7, and $\Omega_{M}$ = 0.3 \citep{2003ApJS..148..175S, 2007ApJS..170..377S}.
At a redshift of $z = 4.055$, 1$^{\prime\prime}$ corresponds to $\sim$7 kpc, and 1 comoving Mpc correspondes to a $\Delta z$ of 0.0014.

\section{OBSERVATIONS \& DATA REDUCTION}
\label{obs}

\subsection{VLA CO(2--1)}
We used the Karl G. Janksy Very Large Array (VLA) to observe the CO(2--1) emission in the GN20 field.
The observations took place in March 2010--April 2011 as part of VLA key project AC974, and they utilized both the D--configuration \citep[28 hours; presented in ][]{2011ApJ...739L..33C} and the higher--resolution B--configuration (96 hours). 
The pointing center was chosen to be 10$^{\prime\prime}$ west of GN20
so that the sources GN20 \citep{2012ApJ...760...11H}, GN20.2a, GN20.2b, a number of nearby LBGs, and BzK--21000 would all fall within the 70\% sensitivity radius of the primary beam.  
The observations were taken in the Q Band, allowing us to simultaneously observe the CO(2--1) emission line (rest frequency $\nu$ $=$ 230.5424 GHz) in the GN20 proto--cluster members ($z = 4.05$), as well as the CO(1--0) emission line in the nearby $z=1.5$ galaxy BzK-21000. 
We centered the two 128 MHz IFs at 45.592 GHz and 45.720 GHz, for a total bandwidth of 246 MHz (taking into account the overlap between IFs). 
At $z = 4.055$, 246 MHz corresponds to a $\Delta z \sim 0.0273$, or $\sim$20 comoving Mpc.
Each IF had 64 channels, corresponding to a spectral resolution of $\sim$13 km s$^{-1}$.  
Observations were taken in full polarization mode.

We reduced the data using standard AIPS tasks.
After accounting for calibration overheads and flagging, the total time on source was approximately 50 hours.
We imaged the data using the AIPS CLEAN algorithm and cleaned down to 1.5$\sigma$ in tight CLEAN boxes around the bright sources (i.e.\ the three SMGs) in the field.
For GN20.2a, the CLEAN box was 0.9$^{\prime\prime}\times0.9^{\prime\prime}$, and for GN20.2b, the CLEAN box was 1.5$^{\prime\prime}\times1.3^{\prime\prime}$.
Briggs weighting with a robust parameter of R $=$ 1.0 resulted in an image cube with an angular resolution of 0.19$^{\prime\prime}$ (1.3 kpc at $z \sim 4.05$) and an rms noise of 74 $\mu$Jy beam$^{-1}$ per 6 MHz (40 km s$^{-1}$) channel.
We extracted spectra for the SMGs using apertures the same size as the CLEAN boxes.
Further details on the data processing can be found in \citet{2012ApJ...760...11H}. 

\begin{figure*}
\centering
\includegraphics[scale=0.65]{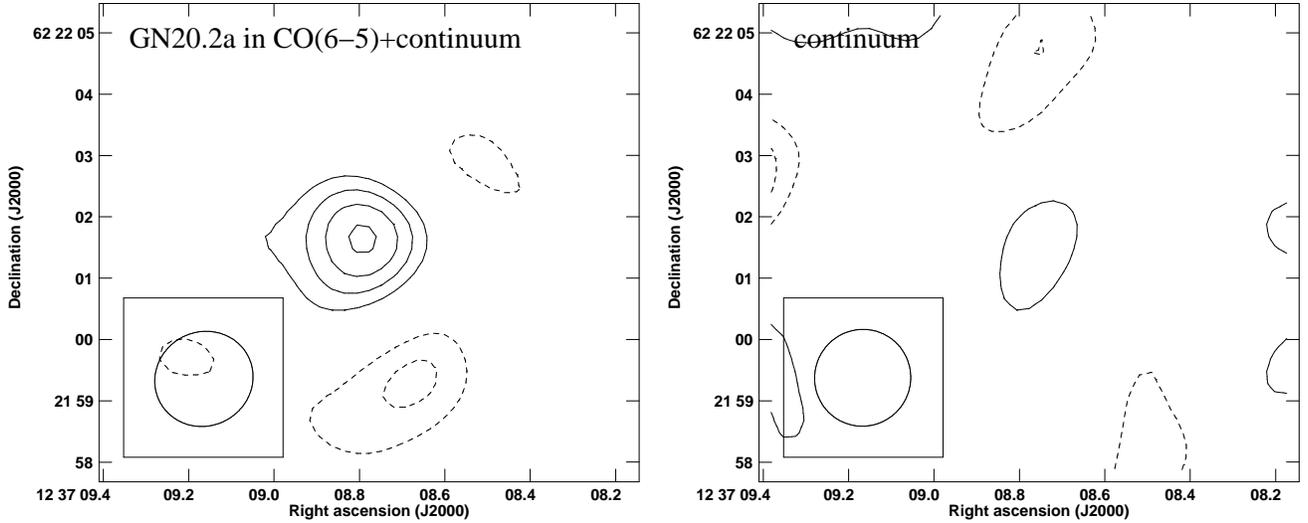}
\caption{PdBI maps of the  CO(6--5)$+$continuum (left) and continuum--only (right) emission for GN20.2a.  The left--hand image was created by averaging over 830 km s$^{-1}$ centered on $z = 4.0512$ (the parameters measured from CO(2--1)). The right--hand image was created by averaging over the line--free emission on either side of the assumed line. Both images have an angular resolution of 1.6$^{\prime\prime}$.  Contours start at $\pm$2$\sigma$ in steps of 1$\sigma$, where $\sigma$ $=$ 390$\mu$Jy beam$^{-1}$ (left) and 380$\mu$Jy beam$^{-1}$ (right). 
}
\label{fig:2a_65maps}
\end{figure*}

\subsection{VLA CO(1--0)}
For the purpose of analyzing the CO excitation of the known SMGs in the field, we made use of several additional datasets examining other CO transitions. 
We used the VLA CO(1--0) data on the GN20 field originally published in \citet{2010ApJ...714.1407C}. 
These D--array data were taken with the old VLA correlator and resulted from a single 50 MHz IF centered at 22.815 GHz. 
The resulting velocity--averaged map had a resolution of 3.7$^{\prime\prime}$ and an rms sensitivity of 30 $\mu$Jy beam$^{-1}$.

\subsection{PdBI CO(4--3)}
\begin{figure}
\centering
\includegraphics[scale=0.65]{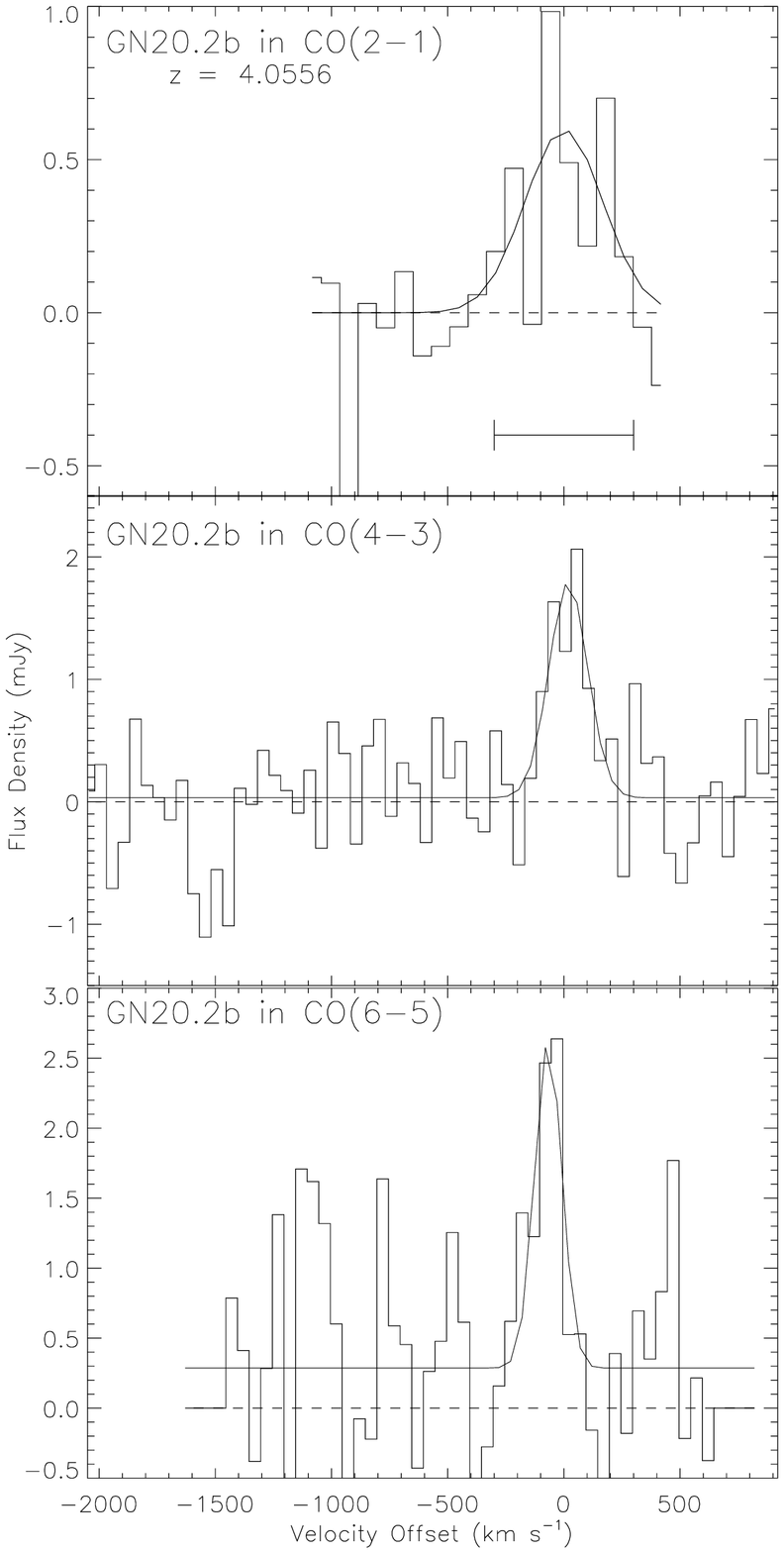}
\caption{CO spectra for GN20.2b from J$=$2--1 (top), 4--3 (middle), and 6--5 (bottom). The corresponding spectral resolutions are 78 km s$^{-1}$, 50 km s$^{-1}$, and 50 km s$^{-1}$, respectively. Gaussian fits to the spectra are shown by the black curves, and the velocity offsets are shown with respect to the derived redshift. 
The bar on the CO(2--1) indicates the velocity range averaged over for the velocity--averaged maps in Figure~\ref{fig:GN20_2_maps}. All spectra have been corrected for the response of the primary beam.}
\label{fig:GN20_2_spectra}
\end{figure}

We made use of the 91 GHz PdBI observations of the GN20 field published in \citet{2009ApJ...694.1517D}. 
These observations were taken in both the B--configuration (1.3$^{\prime\prime}$ synthesized beam) and D--configuration (5.5$^{\prime\prime}$ synthesized beam) and targeted the CO(4--3) emission in the proto--cluster members centered on BzK-21000.
The spectrum for GN20.2a was presented previously in \citet{2009ApJ...694.1517D} and will not be shown again here. 
It was extracted by fitting point sources in the D-- and B--configuration data sets independently and coadding the spectra.
For GN20.2b, we extracted a single--point spectrum from the D--configuration data, 
shown in Figure~\ref{fig:GN20_2_spectra}.
For more information about these data, see \citet{2009ApJ...694.1517D}. 

\subsection{PdBI CO(6--5)}
We used the PdBI to observe the CO(6--5) emission in the GN20 field.
The observations were carried out in May 2008 and January 2009 in the B and D configurations.
The 2mm receivers were tuned to 136.97 GHz to capture both the CO(6--5) emission from the GN20 proto--cluster and the CO(3--2) emission from BzK-21000.
The observations covered a total bandwidth of 1 GHz and were taken in dual polarization mode.

The pointing center for the D--configuration observations was chosen near BzK--21000.
For the B--configuration observations, the pointing center was placed midway between GN20 and its companions GN20.2a and GN20.2b. 
The combined B$+$D data 
were tapered to a resolution of 1.6$^{\prime\prime}$ and imaged, and 
single--point spectra for GN20.2a and GN20.2b were extracted from the image cube.
All images and spectra have been corrected for the response of the PdBI primary beam.
For further details on the analysis of the PdBI data, see \citet{2010ApJ...714.1407C}.

\section{RESULTS}
\label{results}

\subsection{SMG GN20.2a}
\label{gn20_2a_results}

The CO data for the brightest SMG in the field (GN20) have been presented elsewhere \citep{2009ApJ...694.1517D, 2010ApJ...714.1407C, 2011ApJ...739L..33C, 2012ApJ...760...11H} and will not be discussed further. 
We now turn our attention to the other two known SMGs within the field of view of our CO observations: GN20.2a and GN20.2b.
GN20.2a was identified by \citet{2009ApJ...694.1517D} as the primary counterpart to the SCUBA source GN20.2 \citep{2006MNRAS.370.1185P}.
Its CO(2--1) spectrum was shown in \citet{2011ApJ...739L..33C} for the D--configuration data 
and has not changed significantly with the addition of the B--configuration data, presented here for the first time.
A Gaussian fit to the spectrum using the B$+$D--configuration data yields a peak flux density for GN20.2a of 660$\pm$120 $\mu$Jy beam$^{-1}$ and a derived redshift of 4.051$\pm$0.001.
Its FWHM is very broad, at 830$\pm$190 km s$^{-1}$, for a total velocity--integrated CO(2--1) flux density of 0.6$\pm$0.2 Jy km s$^{-1}$.

GN20.2a has a CO(2--1) line luminosity of $L^{\prime}_{\rm CO(2-1)} = 9.7\pm2.9 \times 10^{10}$ K km s$^{-1}$ pc$^{2}$. 
Assuming thermal excitation of CO to extrapolate between CO(2--1) and CO(1--0) 
\citep[as suggested by observations of GN20;][]{2010ApJ...714.1407C, 2012ApJ...760...11H}, 
and using a CO luminosity--to--H$_2$ mass conversion factor of $\alpha_{\rm CO}$ $=$
0.8 M$_{\sun}$ (K km s$^{-1}$ pc$^{2}$)$^{-1}$ as is often used for low--redshift ULIRGs \citep{1998ApJ...507..615D, 2005ARA&A..43..677S, 2006ApJ...640..228T, 2008ApJ...680..246T},
this results in a total molecular gas mass (including Helium) of 7.8$\pm$2.3 $\times$ 10$^{10}$ $\times$ $(\alpha_{\rm CO}/0.8)$ M$_{\sun}$, in agreement with the results from the D--array data alone \citep{2011ApJ...739L..33C}. 

Velocity--averaged CO(2--1) maps for GN20.2a are shown in Figure~\ref{fig:GN20_2_maps} (top row) at three different angular resolutions: 0.77$^{\prime\prime}$ (5.4 kpc; left), 0.38$^{\prime\prime}$ (2.7 kpc; middle), and 0.19$^{\prime\prime}$ (1.3 kpc; right). 
The velocity range covered corresponds to 780 km s$^{-1}$. 
The cross marks the position of the 1.4 GHz counterpart, with the extent of the cross indicating the resolution of those observations \citep[1.7$^{\prime\prime}$;][]{2010ApJS..188..178M}.
Two--dimensional Gaussian fits in the image plane indicate that GN20.2a is marginally--resolved at 0.77$^{\prime\prime}$ resolution and clearly resolved at 0.38$^{\prime\prime}$ resolution.
The deconvolved major axis measured at 0.38$^{\prime\prime}$ resolution is 0.7$^{\prime\prime}\pm0.1^{\prime\prime}$,
and the deconvolved minor axis is 0.4$^{\prime\prime}\pm0.1^{\prime\prime}$ (Gaussian FWHM; i.e.\ $\sim$5 kpc$\times$3 kpc at $z\sim4$).
The total integrated flux density measured from the two--dimensional Gaussian fit at 0.38$^{\prime\prime}$ resolution is consistent with the lower--resolution observations.
At the highest--resolution (0.19$^{\prime\prime}$), there is one significant ($>$4$\sigma$) component
situated almost on top of the radio position. 
This component is marginally--resolved and contains $\sim$half of the total flux density of the source.

The CO(4--3) spectrum for GN20.2a was presented in \citet{2009ApJ...694.1517D}.
They reported a 7$\sigma$ detection in the PdBI B$+$D--configuration data with a velocity--integrated flux density of $I_{\rm CO} = 0.9\pm0.3$ Jy km s$^{-1}$. 
We will use this value in our spectral line energy distribution modeling in Section~\ref{analysis}.

The CO(6--5) spectrum of GN20.2a (not shown) is low S/N, and we therefore fixed the redshift and FWHM to the CO(2--1) values when fitting a Gaussian model.
A formal fit to the spectrum gives a peak of 1.6$\pm$0.4 mJy and a continuum level of 0.7$\pm$0.2 mJy. 
The velocity--integrated CO(6--5) flux density is 1.4$\pm$0.5 Jy km s$^{-1}$. 
Velocity--averaged maps are shown in Figure~\ref{fig:2a_65maps} for the CO(6--5)$+$continuum (left) and continuum--only (right) emission.
The CO(6--5)$+$continuum map was created by averaging over 830 km s$^{-1}$ (the FWHM measured from CO(2--1), centered on $z = 4.0512$ (the measured redshift of CO(2--1)).
The continuum--only image was restricted to only a few line--free channels on either side of the assumed line, 
and the measured flux density is consistent with the spectral fit within the error bars.
Even though the spectrum is low S/N, these images demonstrate that the CO(6--5) line emission is significant when averaged over the line FWHM.

Finally, we examined the CO(1--0) data on the GN20 field. 
Neither GN20.2a nor GN20.2b were detected individually, though there is possible evidence for an increase in flux density in the vicinity of the sources.
Note that the 50 MHz bandpass is quite narrow, and the velocity--averaged map only covers a portion of the linewidth derived for GN20.2a from the CO(2--1) observations. 
Therefore, taking GN20.2a as a non--detection, and scaling up the upper limit derived from the rms sensitivity by assuming a linewidth of 830 km s$^{-1}$, 
we derive a (corrected) 5$\sigma$ upper limit of I$_{\rm CO}$ $<$ 0.15 Jy km s$^{-1}$.
This corresponds to $L^{\prime}_{\rm CO(1-0)} < 10.0 \times 10^{10}$ K km s$^{-1}$ pc$^{2}$
and M(H$_{2}$) $<$ 8.0 $\times$ 10$^{10}$ $\times$ $(\alpha_{\rm CO}/0.8)$ M$_{\sun}$, consistent 
with the assumption of thermalization.

\begin{figure*}
\centering
\includegraphics[scale=0.65]{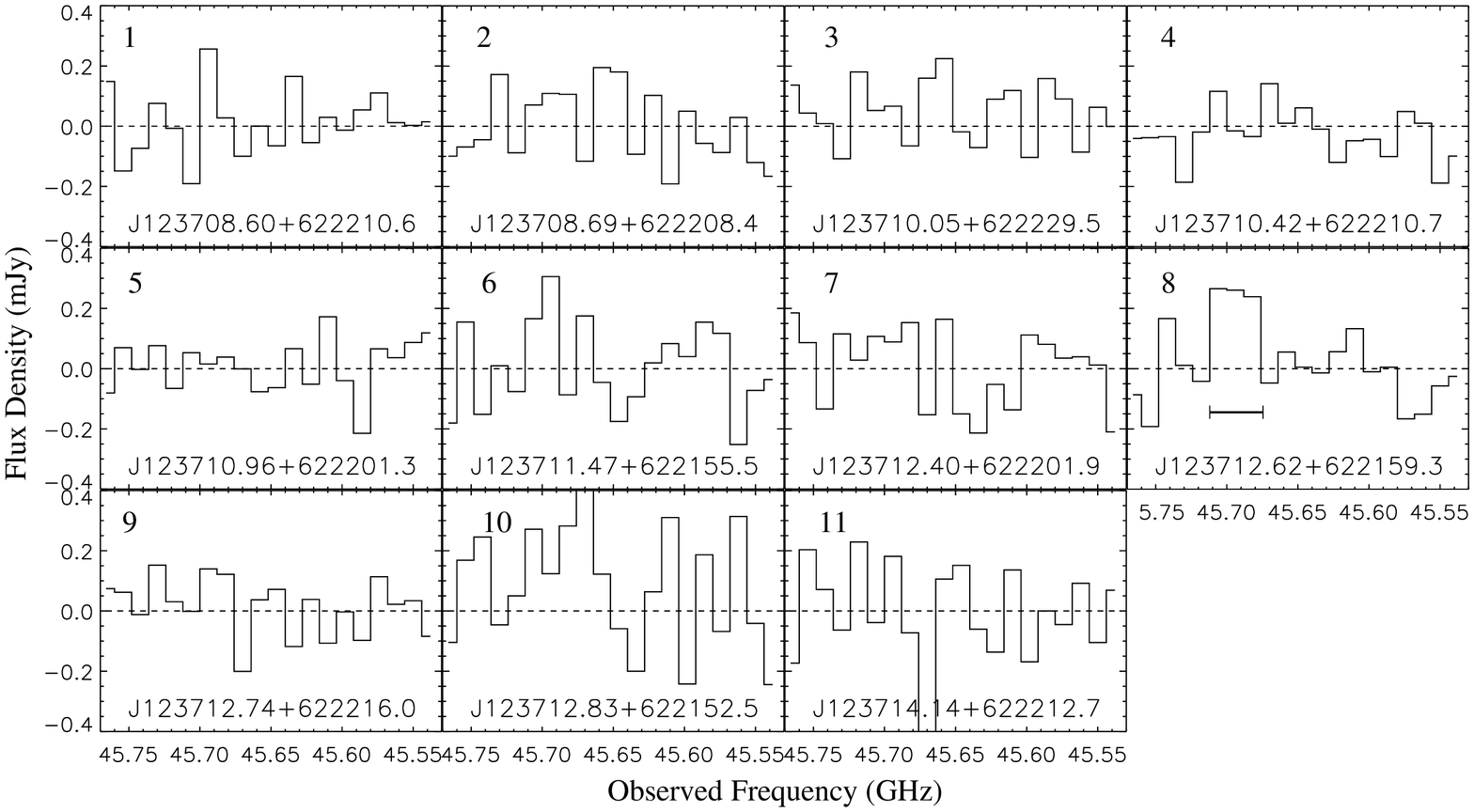}
\caption{$B$--band dropout spectra at an angular resolution of 1.4$^{\prime\prime}$ and a spectral resolution of 80 km s$^{-1}$. These spectra have not been corrected for the primary beam response. Source \#8 shows evidence for a possible emission line. Averaging over the 36 MHz range indicated on the 
spectrum produces the velocity-averaged CO contours shown in Figure~\ref{fig:bdrop11+HST}. See Section~\ref{lbgs} for details.}
\label{fig:bdrops_130kL}
\end{figure*}

\subsection{SMG GN20.2b}
\label{gn20_2b_results}

The CO(2--1) spectrum for GN20.2b is shown in Figure~\ref{fig:GN20_2_spectra}, smoothed to 78 km s$^{-1}$ (12 MHz) per channel.  
A Gaussian fit to the spectrum is shown by the black curve, and the velocity offset is shown with respect to the derived redshift.
The fit has a peak flux density of 600$\pm$270 $\mu$Jy beam$^{-1}$ and a derived redshift of 4.056$\pm$0.001.
The FWHM of GN20.2b is significantly smaller than GN20.2a, at 400$\pm$210 km s$^{-1}$.  
The continuum level was held fixed at zero due to the small amount of line-free emission observable.
The gaussian fit implies a total velocity--integrated CO(2--1) flux density of 0.3$\pm$0.2 Jy km s$^{-1}$
and a line luminosity of $L^{\prime}_{\rm CO(2-1)} = 4.2\pm2.9 \times 10^{10}$ K km s$^{-1}$ pc$^{2}$.
Again assuming thermal excitation of CO and a CO luminosity--to--H$_2$ mass conversion factor of 0.8 M$_{\sun}$ (K km s$^{-1}$ pc$^{2}$)$^{-1}$,
we measure a molecular gas mass of $3.4\pm2.3 \times 10^{10} \times (\alpha_{\rm CO}/0.8)$ M$_{\sun}$, consistent with the estimated value from the D--array data alone \citep{2011ApJ...739L..33C}.

The velocity--averaged CO(2--1) maps for GN20.2b are shown in Figure~\ref{fig:GN20_2_maps} (bottom row), again at three different angular resolutions: 0.77$^{\prime\prime}$ (5.4 kpc; left), 0.38$^{\prime\prime}$ (2.7 kpc; middle), and 0.19$^{\prime\prime}$ (1.3 kpc; right). 
The velocity range covered is indicated in Figure~\ref{fig:GN20_2_spectra} and corresponds to 
470 km s$^{-1}$. 
GN20.2b is already clearly resolved at the lowest resolution shown here (0.77$^{\prime\prime}$).
Its deconvolved size is (1.1$^{\prime\prime}\pm0.4^{\prime\prime}$)$\times$(0.7$^{\prime\prime}\pm0.4^{\prime\prime}$) (Gaussian FWHM; i.e.\ $\sim$8 kpc$\times$5 kpc). 
The 0.38$^{\prime\prime}$ resolution map shows one $>$4$\sigma$ component which is resolved and has an integrated flux density consistent with the total flux density of the source at lower resolution but with substantial error bars.
GN20.2b is entirely resolved out in the highest--resolution (0.19$^{\prime\prime}$) map.

The CO(4--3) and CO(6--5) spectra for GN20.2b are also shown in Figure~\ref{fig:GN20_2_spectra}.
For the CO(4--3), we detect no continuum emission, and a peak flux density of 1.8$\pm$0.4 mJy. 
The velocity--integrated CO(4--3) flux density is 0.4$\pm$0.1 Jy km s$^{-1}$. 
For CO(6--5), the peak flux density is 2.3$\pm$0.6 mJy, and
the velocity--integrated CO(6--5) flux density is 0.3$\pm$0.1 Jy km s$^{-1}$.
The centroid of the line emission appears slightly offset from the CO(2--1) value used in the Gaussian fit, but we note that the spectrum is low S/N.
We detect a continuum level of 0.3$\pm$0.1 mJy which, combined with the continuum level derived for GN20.2a at the same frequency, is consistent with the extrapolation of the 850$\mu$m SCUBA flux density of GN20.2 to 136 GHz (0.94 mJy, assuming an Arp220--like spectrum) and is divided over the two sources in the same fractions as estimated for the submillimeter emission \citep{2009ApJ...694.1517D}.

Lastly, as with GN20.2a, GN20.2b is undetected in the (single--channel) CO(1--0) data.
The 50 MHz bandwidth of the channel only covers a portion of the expected linewidth, with the line center close ($\sim$90 km s$^{-1}$) to the edge of the band.
Correcting for this effect, we derive a 5$\sigma$ upper limit for CO(1--0) in GN20.2b of I$_{\rm CO}$ $<$ 0.12 Jy km s$^{-1}$.
This corresponds to $L^{\prime}_{\rm CO(1-0)} < 8.0 \times 10^{10}$ K km s$^{-1}$ pc$^{2}$
and M(H$_{2}$) $<$ 6.4 $\times$ 10$^{10}$ $\times$ $(\alpha_{\rm CO}/0.8)$ M$_{\sun}$, again consistent with thermalization within the uncertainties.

\begin{figure}
\centering
\includegraphics[scale=0.5]{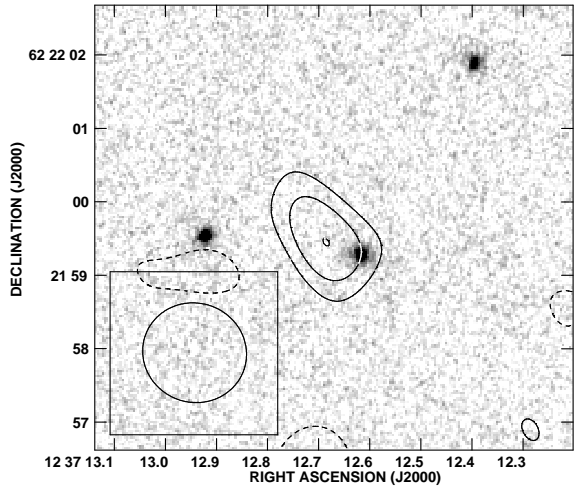}
\caption{CO(2--1) contours from the velocity--averaged 45 GHz VLA data at 1.4$^{\prime\prime}$ resolution overlaid on the deep HST$+$ACS z850--band image for LBG \#8 from Figure~\ref{fig:bdrops_130kL}. The VLA map has been corrected for the primary beam response and has an rms noise of 82 $\mu$Jy beam$^{-1}$. Contours are given in steps of 1$\sigma$ starting at $\pm$2$\sigma$. The velocity--averaged map was created by averaging over 36 MHz centered on the tentative emission line in Figure~\ref{fig:bdrops_130kL} at 45.694 GHz. If the emission line is real and is from CO(2--1), then the corresponding redshift of LBG \#8 is 4.0452$\pm$0.0004. }
\label{fig:bdrop11+HST}
\end{figure}

\subsection{CO in proto--cluster LBGs}
\label{lbgs}

The GN20 field contains an overdensity of $z\sim4$ Lyman break galaxies ($B$--band dropouts), with 14 such sources within just 25$^{\prime\prime}$ of GN20 (Figure~\ref{fig:field}).
The presence of these galaxies in the field allows us to study the molecular gas content in lower star formation rate galaxies in the early universe.
We therefore conducted a targeted search for their CO emission using the deep CO(2--1) VLA data.
We used the data cube tapered to 1.4$^{\prime\prime}$ resolution, corresponding to $\sim$10 kpc at the proto--cluster redshift.
This ensured that we would be probing the total molecular gas content in each galaxy with single--point spectra.
We extracted spectra at the positions of the HST$+$ACS sources, and we show them in Figure~\ref{fig:bdrops_130kL}.
Two of the LBGs are the likely counterparts to the known sources GN20 and GN20.2a, and a third is only $\sim$0.7$^{\prime\prime}$ from GN20.2a and therefore confused at this resolution -- these three LBGs are not included in Figure~\ref{fig:bdrops_130kL}.
(Note that Daddi et al.\ 2010 has suggested that the two LBGs near GN20.2a may be in the process of merging -- see Section~\ref{localization}).
The LBG spectra are consistent with noise, except for a possible line in \#8 at 45.694 GHz. 
A Gaussian fit to the line gives a FWHM of 170$\pm$65 km s$^{-1}$.
If the detection is real and is due to CO(2--1) line emission, then the corresponding redshift is 4.0452$\pm$0.0004. 
Contours from a velocity--averaged map (over the 36 MHz specified in Figure~\ref{fig:bdrops_130kL}) are shown overlaid on a zoomed--in region of the deep HST$+$ACS z850--band cutout for this LBG in Figure~\ref{fig:bdrop11+HST}, and the LBG is indicated in Figure~\ref{fig:field} as a diamond symbol.
We will discuss the implications of these results in Section~\ref{analysis}.

\subsection{Blind search for CO emission--line sources}
\label{serch}

As a final step toward extracting all possible information from the deep VLA CO(2--1) data cube,
we conducted a blind search for emission--line sources in the field using the AIPS task SERCH. 
SERCH looks for emission in an input data cube based on a range of expected linewidths and a S/N threshold.
We used an input image cube with a spectral resolution of 40 km s$^{-1}$ and a $uv$--taper of 130k$\lambda$, corresponding to a resolution of 1.4$^{\prime\prime}$, or 10 kpc for sources in the $z=4.05$ GN20 proto--cluster. 
This resolution ensured that we would not resolve--out potentially extended source candidates.
To 
enable a search with uniform signal--to--noise levels in the spatial domain, 
the cube was not corrected for the response of the VLA primary beam (although any CO estimates resulting from the search have been corrected).

\begin{figure}
\centering
\includegraphics[scale=0.5]{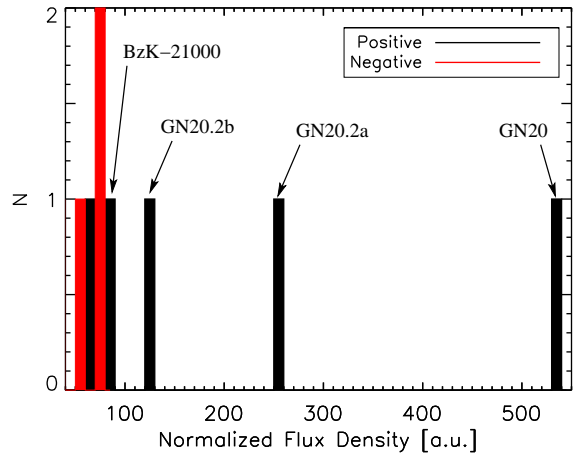}
\caption{Histogram of emission line intensity (in arbitrary units) for the positive (black) and negative (red) source candidates recovered using SERCH on the VLA data cube tapered to 1.4$^{\prime\prime}$ ($\sim$10 kpc at $z \sim 4.05$) resolution. The intensities have not been corrected for the primary beam response.}
\label{fig:serch}
\end{figure}

Within the SERCH task, we used a smoothing kernel of 4--12 channels, corresponding to velocity widths of 160--480 km s$^{-1}$.
Using a S/N threshold of 4.5, 
SERCH identified five positive peaks and three negative peaks within the 50\% sensitivity threshold of the primary beam.
The positive peaks (our source candidate list) include the four sources which have been detected previously in CO emission: the SMGs GN20, GN20.2a, and GN20.2b in CO(2--1), 
and the $z\sim1.5$ galaxy BzK--21000 in CO(1--0) \citep[][and Figure~\ref{fig:field} of this work]{2009ApJ...694.1517D,2010ApJ...713..686D}.
Note that with these small number statistics, we only expect 
2$\pm\sqrt{3}$
sources to be real detections.
We therefore used the strengths of the lines to glean further information on the possible detections.
We fit all of the peaks with Gaussians, and we plot a histogram of both the positive and negative peaks as a function of line intensity (based on the fits).
The result is shown in Figure~\ref{fig:serch}, where the positive peaks are shown in black and the negative peaks are shown in red. 
The three known $z\sim4$ SMGs are the strongest sources overall, with intensites greater than the strongest noise peaks.
The galaxy BzK--21000 has a similar CO luminosity to the strongest negative peaks in the cube, implying its detection is marginal even at this resolution.
The one unkown source candidate falls in the noise and is likely spurious.
We therefore do not find any additional source candidates with this search.

The 1.4$^{\prime\prime}$ taper applied to the above data cube came at the cost of decreased sensitivity, as the lowest noise is achieved in a data cube with an untapered resolution of 0.19$^{\prime\prime}$.
However, this resolution corresponds to only 1.3 kpc at $z = 4.05$, making it difficult to detect more extended sources whose emission is spread out over many beam areas.  
Therefore, in order to increase our sensitivity to faint sources without running the risk of resolving them out, 
we repeated the search on a cube with a $uv$--taper of 600k$\lambda$, or a resolution of 0.38$^{\prime\prime}$/2.6 kpc for sources in the $z=4.05$ GN20 proto--cluster. 
This small amount of taper allowed us to achieve an rms noise per channel that is only marginally higher than the best case value, while also allowing a factor of $\sim$4 increase in beam area.
Assuming a linewidth of 300 km s$^{-1}$, this corresponds to a 5$\sigma$ limiting CO luminosity of
$L^{\prime}_{\rm CO(2-1)} = 8 \times 10^{9}$ K km s$^{-1}$ pc$^{2}$ at $z\sim4.05$.

\begin{figure*}
\centering
\includegraphics[scale=0.7]{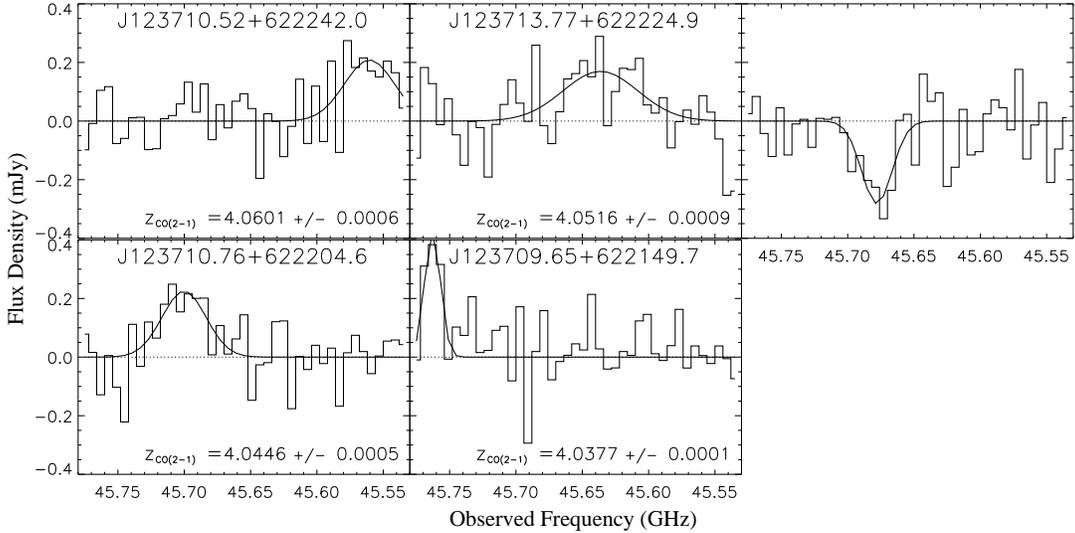}
\caption{Spectra for four of the six positive peaks and the negative peak recovered by SERCH from the VLA data cube tapered to 0.38$^{\prime\prime}$ resolution. Not shown are the positive detections for GN20 and GN20.2a. These spectra have not been corrected for the primary beam response. 
The redshifts listed for the four source candidates assume the emission line is real and due to CO(2--1).}
\label{fig:serch_spectra_600kL}
\end{figure*}

Using the same range of smoothing kernels and a S/N threshold of 5.1 (determined to produce 
only one negative peak),
SERCH identified six positive peaks and 
one negative peak within the 50\% sensitivity threshold of the primary beam.
Two of the positive peaks correspond to GN20 and GN20.2a.
Spectra for the four remaining positive peaks (source candidates) and 
one negative peak are shown in Figure~\ref{fig:serch_spectra_600kL}, 
with the coordinates and possible redshifts of the positive peaks indicated.
The redshifts assume the emission line is real and due to CO(2--1).
Within the limited statistics, we therefore expect $\sim$5$\pm$1 real sources (including GN20 and GN20.2a).

\begin{figure}
\centering
\includegraphics[scale=0.55]{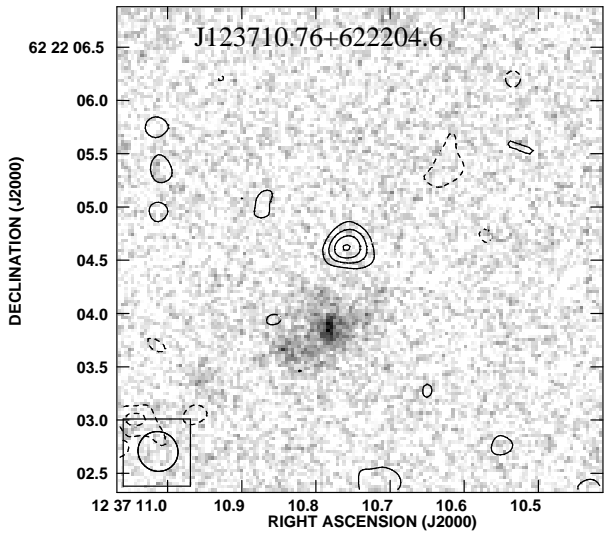}
\caption{Emission line candidate J123710.76+622204.6 from a blind search of the deep 45 GHz VLA data tapered to 0.38$^{\prime\prime}$ resolution. Contours show the velocity--averaged 45 GHz map and start at $\pm$2$\sigma$, where $\sigma = 43 \mu$Jy beam$^{-1}$. The background image is a cutout from the HST$+$ACS 850z--band imaging.}
\label{fig:PosSrc+HST}
\end{figure}

The positions of the four previously unnamed emission line candidates are indicated in Figure~\ref{fig:field} with star symbols. (GN20 and GN20.2a, which were also recovered in this search, are already indicated with large crosses.) Only one of the previously unnamed source candidates has a possible optical counterpart (J123710.76+622204.6, circled star in Figure~\ref{fig:field}). Figure~\ref{fig:PosSrc+HST} show its velocity--averaged VLA contours overlaid on a $\sim$4$^{\prime}$$\times$4$^{\prime}$ cutout from the deep HST$+$ACS 850z--band coverage of GOODS-N. The 45 GHz emission line candidate lies within 1$^{\prime\prime}$ of an optical galaxy. There is an IRAC source coincident (within 0.078$^{\prime\prime}$) with the optical galaxy, and the corresponding photometric-redshift is 1.34. This could indicate that the 45 GHz emission line, if real, is CO(1--0). Its redshift (as measured from our spectrum) would then be $z\sim1.52$, consistent with that of the possible counterpart given typical photometric redshift uncertainties. The positional offset (0.9$^{\prime\prime}$) between the IRAC and CO positions implies a small but non--negligible chance that the two are related ($\sim$5\% probability). Therefore, as an additional test, we checked whether the CO--derived properties would make sense in light of the possible counterpart. Assuming that the emission line is real, and using the relation between CO luminosity and IR luminosity determined for normal galaxies \citep[i.e.\ local spirals and $z\sim1.5$ BzK galaxies;][]{2010ApJ...713..686D}, we derived an IR luminosity of $L_{\rm IR}$ $=$ 5 $\times$ 10$^{11}$ L$_{\odot}$ and a SFR \citep[using the relation from ][adapted to a Chabrier IMF]{1998ARA&A..36..189K} of $\sim$50 M$_{\odot}$ yr$^{-1}$. On the other hand, using the $B$--band luminosity of the ACS source to estimate the rest--frame ultraviolet flux, and estimating the extinction and SFR using equations appropriate for $z\sim1.5$ BzK galaxies gives a SFR of only 8 M$_{\odot}$ yr$^{-1}$. The UV--derived SFR is therefore $\sim$6$\times$ lower than expected from the observed CO emission. If it were a starburst, on the other hand, the IR luminosity implied by the CO would be $\sim$3$\times$ larger (using the ULIRG $L_{\rm CO}/L_{\rm IR}$ ratio), but the UV--derived properties would not be expected to match in the presence of a significant amount of dust obscuration. Such dust obscuration could also perhaps explain the offset between the CO and ACS counterparts. Therefore, if this CO emission line is real, and if we have identified its counterpart, then the source must be a starburst.


\section{ANALYSIS}
\label{analysis}

\subsection{GN20.2a \& GN20.2b} 
\label{GN20_2_anal}

\subsubsection{CO sizes and gas surface densities} 
\label{COsize}

The high--resolution CO(2--1) data allow us to measure (projected) sizes for GN20.2a and GN20.2b of 5 kpc$\times$3 kpc (GN20.2a), and 8 kpc$\times$5 kpc (GN20.2b).
For comparison, previous CO size measurements for SMGs have reported typical diameters of $<$4--5 kpc 
\citep{2006ApJ...640..228T, 2008ApJ...680..246T, 2010ApJ...724..233E, 2013MNRAS.429.3047B}.
However, the majority of these measurements utilize higher--J (J$\geq$3) CO transitions, which may not measure the full extent of the cold gas reservoir.
Some observations in CO(1--0), for example, find very extended gas reservoirs, with sizes of $\sim$15 kpc \citep[e.g.][]{2010MNRAS.404..198I,2011MNRAS.412.1913I, 2011ApJ...733L..11R, 2011ApJ...739L..31R}.
The compactness we measure in GN20.2a in CO(2--1) is closer to the sizes typically measured in higher--J transitions than to the very large sizes seen in some SMGs in CO(1--0).
If the extended CO(1--0) gas reservoirs are representative of the SMG population as a whole, then either 
GN20.2a is
more compact than most, or the CO(2--1) emission also underestimates the full extent of the gas reservoir. 
Indeed, a recent study by \citet{2013MNRAS.429.3047B} reports a small transition--dependent effect, with SMGs observed in CO(3--2) having slightly larger sizes ($\sim$6 kpc) than those observed in CO(4--3) ($\sim$4 kpc; though they caution that their sample size is small),
so it is possible that this effect also exists between CO(1--0) and CO(2--1).

In GN20.2b, on the other hand, we see a more extended gas reservoir, with a size roughly double that of typical CO(3--2)/CO(4--3) sizes. 
For comparison, note that although \citet{2012ApJ...760...11H} report a size of 14 kpc for GN20 in CO(2--1), 
this was only after applying a careful masking procedure to recover the extended, diffuse emission. 
In the regular, velocity--averaged CO(2--1) map, GN20 has an apparent size of only $\sim$7 kpc \citep{2012ApJ...760...11H}, 
similar to what we see in GN20.2b.

From these estimates, we derive gas mass surface densities for the SMGs,
making the assumption that the FWHM sizes measured contain half the total amount of gas.
GN20.2b has an average gas surface density of 530$\times$(sin \textit{i}) $(\alpha_{\rm CO}/0.8)$ M$_{\sun}$ pc$^{-2}$, where \textit{i} is the unknown inclination.
The one significant ($>$4$\sigma$) component in the 2.7 kpc resolution map of GN20.2b has a surface density of 1700$\times$(sin \textit{i}) $(\alpha_{\rm CO}/0.8)$ M$_{\sun}$ pc$^{-2}$.
For the more compact SMG GN20.2a, the average gas surface density is $\sim$3900$\times$(sin \textit{i}) $(\alpha_{\rm CO}/0.8)$ M$_{\sun}$ pc$^{-2}$, where its inclination \textit{i} is again unknown.
GN20.2a has one significant ($>$4$\sigma$) component in the 0.19$^{\prime\prime}$/1.3 kpc resolution map which has a surface density of $\sim$12,700$\times$(sin \textit{i}) $(\alpha_{\rm CO}/0.8)$ M$_{\sun}$ pc$^{-2}$.

Comparably high gas surface densities have been observed previously in certain SMGs, ULIRGs, and high--redshift quasars \citep{2008ApJS..178..189W, 2008ApJ...686L...9R, 2010ApJ...724..233E, 2011ApJ...742...11S}.
On the other hand, local spirals have typical gas surface densities of only 1--100 M$_{\sun}$ pc$^{-2}$ \citep[e.g.,][]{2008AJ....136.2846B}, and giant molecular clouds (GMCs) reach values of $\sim$200 M$_{\sun}$ pc$^{-2}$.
At slightly higher redshift ($z=1-2.3$), normal galaxies have average gas surface densities of 50--2500 M$_{\sun}$ pc$^{-2}$ \citep{2010Natur.463..781T}, comparable to the average density we observe in GN20.2b.
The maximum surface density of GN20.2a is still well outside this range, however, and it may be that the only way to achieve such high surface densities is through the tidal torquing resulting from the final stages of a major merger \citep{2008ApJ...680..246T}.

\subsubsection{Star formation rate density} 
\label{sfe}

GN20.2a has an estimated SFR from \citet{2009ApJ...694.1517D}, allowing us to calculate several other parameters of interest for this SMG.
Based on its IR luminosity, GN20.2a has a SFR of $\sim$1600 M$_{\sun}$ yr$^{-1}$ \citep{2009ApJ...694.1517D}.
Taking its estimated size of $\sim$5 kpc $\times$ 3 kpc, 
and assuming that half the star formation occurs within the half--light radius,
this corresponds to an average SFR density of $\sim$80$\times$(sin \textit{i}) M$_{\sun}$ yr$^{-1}$ kpc$^{-2}$.
At this resolution, the SFR density is well below the theoretical value for Eddington--limited maximal starbursts \citep[1000 M$_{\sun}$ yr$^{-1}$ kpc$^{-2}$;][]{2005ApJ...630..167T}.
Its gas consumption timescale (M(H$_2$)/SFR) is $\sim$50 ($\alpha_{\rm CO}$/0.8) Myr, 
an order of magnitude shorter than normal star forming galaxies at $z\sim2$ \citep[e.g.,][]{2010Natur.463..781T}.

\subsubsection{Dynamical masses and the CO--to--H$_2$ conversion factor}
\label{Mdyn}

We estimate the dynamical masses of GN20.2a and GN20.2b using 
the measured FWHM velocity of the CO(2--1) lines and the observed semi--major axes.
We follow \citet{2008ApJ...680..246T} and take the average of two different estimators: the usual isotropic virial estimator
\begin{equation}
M_{\rm dyn} = \frac{5\sigma^2R}{G}
\end{equation}
(where $\sigma$ is the one--dimensional velocity dispersion, $R$ is the semi--major axis, and $G$ is the gravitational constant), and the global rotating disk estimator, corrected for $\langle \rm{sin}^2(i) \rangle = 2/3$ in mass:
\begin{equation}
M_{\rm dyn} = 6 \times 10^4 \Delta\nu^{2}_{\rm FWHM}R
\end{equation}
where $\Delta\nu^{2}_{\rm FWHM}$ is the line width FWHM, and $R$ is again the semi--major axis.
Taking the average of these two estimators, the implied dynamical mass of GN20.2a within $R\sim2.5$ kpc is (2.2$\pm$0.9) $\times$ 10$^{11}$ M$_{\odot}$,
and the dynamical mass of GN20.2b is 6$\pm^{7}_{6}$ $\times$ 10$^{10}$ M$_{\odot}$  within $R\sim4$ kpc.
The large error bar for GN20.2b reflects the large uncertainty in its fitted linewidth, which comes into the above equations squared.

\begin{figure*}
\centering
\includegraphics[scale=0.6]{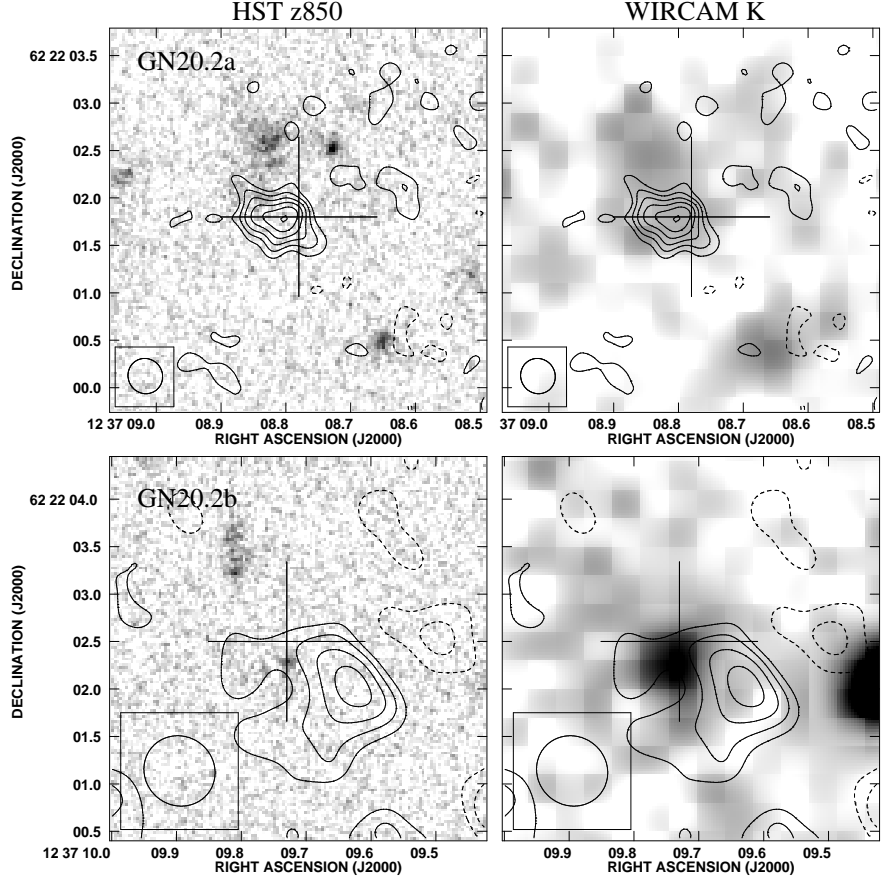}
\caption{VLA CO(2--1) velocity--averaged contours for GN20.2a (top) and GN20.2b (bottom) overlaid on the HST$+$ACS 850$z$--band image (left) and the WIRCAM K--band image (right). The CO(2--1) contours are shown at 0.38$^{\prime\prime}$/2.7 kpc resolution for GN20.2a, and at 0.77$^{\prime\prime}$/5.4 kpc resolution for gN20.2b. The cross is the same as in Figure~\ref{fig:GN20_2_maps}. The CO maps have been corrected for the response of the primary beam. The contrast in the HST maps is the same for GN20.2a and GN20.2b (and the same is true for the WIRCAM K--band images of the galaxies).}
\label{fig:GN20_2_multiwav}
\end{figure*}

\begin{deluxetable*}{ l l l l l}
\tabletypesize{\small}
\tablewidth{0pt}
\tablecaption{GN20.2a/b Derived Values \label{tab-1}}
\tablehead{
\colhead{\textbf{Source}} & \colhead{\textbf{$I_{\rm CO(1-0)}$}} & \colhead{\textbf{$I_{\rm CO(2-1)}$}} & \colhead{\textbf{$I_{\rm CO(4-3)}$}} & \colhead{\textbf{$I_{\rm CO(6-5)}$}} \\
\colhead{} & \colhead{[Jy km s$^{-1}$]} & \colhead{[Jy km s$^{-1}$]} & \colhead{[Jy km s$^{-1}$]} & \colhead{[Jy km s$^{-1}$]} }
\startdata
GN20.2a 	& $<$0.15$^{a}$ 	& 0.6$\pm$0.2	& 0.9$\pm$0.3	& 1.4$\pm$0.5 \\
GN20.2b		& $<$0.12$^{a}$		& 0.3$\pm$0.2	& 0.4$\pm$0.1	& 0.3$\pm$0.1 
\enddata
\tablenotetext{a}{5$\sigma$ upper limit.}
\end{deluxetable*}

The derived dynamical masses can be used to constrain the CO--to--H$_2$ conversion factor for these systems.
As the dynamical masses were computed within the semi--major axes (i.e., HWHM), we assume \textit{half} the stellar and gas masses when backing out $\alpha_{\rm CO}$ values \citep{2010ApJ...713..686D}.
If GN20.2a were composed entirely of molecular gas, the implied CO--to--H$_2$ conversion factor would be $\alpha_{\rm CO}$ $=$ 4.5$\pm$2.2 M$_{\sun}$ (K km s$^{-1}$ pc$^{2}$)$^{-1}$ (including Helium).
Clearly this is an extreme assumption, but we make it to illustrate that the derived value
is 
then consistent with the 
local value of $\alpha_{\rm CO}$ $=$ 4.3 M$_{\sun}$ (K km s$^{-1}$ pc$^{2}$)$^{-1}$ (including Helium), which has been well--established for the Milky Way, nearby star--forming galaxies, and even dense star--forming clumps of lower--mass/metallicity galaxies \citep{1996A&A...308L..21S, 2001ApJ...547..792D, 2005Sci...307.1292G, 2008ApJ...686..948B, 2010ApJ...710..133A, 2011ApJ...737...12L}.
Including the stellar mass and dark matter content will 
decrease this factor.
\citet{2009ApJ...694.1517D} previously estimated the stellar mass of GN20.2a to be 5 $\times$ 10$^{10}$ M$_{\odot}$ from SED fitting to the ACS through IRAC photometry.
If we take this stellar mass,
assuming an uncertainty of 
0.3dex due to the extreme obscuration in the UV and the systematic uncertainties in star formation histories,
and if we assume a dark matter content of 25\% \citep[][and references therein]{2010ApJ...713..686D}
we find $\alpha_{\rm CO}$ $=$ 2.9$\pm^{1.7}_{1.6}$ M$_{\sun}$ (K km s$^{-1}$ pc$^{2}$)$^{-1}$.
As the relation of the presumed optical counterpart to the CO--emitting region is unclear (see Section~\ref{localization}), if we instead simply assume 25\% dark matter, with the remaining mass split equally between gas and stars, we arrive at a conversion factor of $\alpha_{\rm CO}$ $=$ 1.7$\pm0.8$ M$_{\sun}$ (K km s$^{-1}$ pc$^{2}$)$^{-1}$.
The data may therefore be more consistent with a lower conversion factor,
although it is difficult to draw any firm conclusions given all of the uncercainties that went into this calculation.

The error on GN20.2b's dynamical mass estimate already indicates that the constraints on $\alpha_{\rm CO}$ will be weak, but we go through the analysis here to be thorough.
Based on its measured CO luminosity and estimated dynamical mass, we derive a CO--to--H$_2$ conversion factor of $\alpha_{\rm CO}$ $=$ (3.0$\pm^{4.0}_{3.0}$) M$_{\sun}$ (K km s$^{-1}$ pc$^{2}$)$^{-1}$ \textit{if} we assume that GN20.2b is composed entirely of molecular gas.
We have no stellar mass estimate for GN20.2b, but if we assume 25\% dark matter, with the remaining mass split equally between gas and stars, we derive a conversion factor of
$\alpha_{\rm CO}$ $=$ (1.1$\pm^{1.5}_{1.1}$) M$_{\sun}$ (K km s$^{-1}$ pc$^{2}$)$^{-1}$.
This value supports our previous assumption of a low, ULIRG--like conversion factor, 
and it agrees with estimates derived for nearby GN20 by \citet{2012ApJ...760...11H} and \citet{2011ApJ...740L..15M}. 

\subsubsection{Localization of Counterparts}
\label{localization}

The CO(2--1) velocity--averaged contours for GN20.2a and GN20.2b are shown overlaid on a selection of multiwavelength data in Figure~\ref{fig:GN20_2_multiwav}.
The contours are shown at 0.38$^{\prime\prime}$/2.7 kpc resolution for GN20.2a, and at 0.77$^{\prime\prime}$/5.4 kpc resolution for GN20.2b.
The left panel shows the HST$+$ACS 850z--band image, and the right panel shows the corresponding WIRCAM K--band image. 
For GN20.2a (top panels), 
the diffuse HST galaxy to the northeast of the cross was identified by \citet{2009ApJ...694.1517D} as the likely optical counterpart to the CO emission.
Based on Keck$+$DEIMOS spectroscopy, its redshift is $z = 4.059\pm0.007$, consistent with the redshift we derive from the CO(2--1) emission. 

For GN20.2b (bottom panels),
the likely optical counterpart is the very faint galaxy ($z_{\rm AB} = 27.34$) near the cross at the center of the 850z--band image. 
Using a radio--IR photometric redshift technique, and assuming the emission is dominated by star formation, \citet{2009ApJ...694.1517D} constrained its redshift to $z > 3.2$ at the 99\% confidence level.  
Their tentative CO detection meant that they could not rule out a radio--loud AGN at low--redshift, but their evidence was also consistent with a highly star--forming galaxy at high redshift. 
Our CO(2--1) detection is only $\sim$5.5$\sigma$ (averaged over the linewidth) but is consistent with a redshift of $z \sim$ 4.05, strengthening the case put forward in \citet{2009ApJ...694.1517D} based on a weak detection of CO(4--3). 

One thing that is now very clear from the sub--arcsecond resolution CO(2--1) imaging is the significant offset
between the CO emission and the HST$+$ACS counterparts (particularly for GN20.2a, but present for GN20.2b as well).
Such large offsets are not unusual for SMGs and may indicate the presence of a substantial dust screen roughly coincident with the CO emission and blocking the majority of the UV/optical emission from the galaxy. 
Such a scenario has been proposed for the SMG GN20 to explain its large offset \citep{2012ApJ...760...11H}. 
For this scenario to be viable, the dust screen must extend over 10s of kpc in order to block most of the emission from the disk,
and/or the inherent UV/optical morphology must be irregular/asymmetric.
Alternately, it may be that the optical ``counterparts" are distinct galaxies, either unrelated or in the process of merging with the dusty starburst galaxy (traced by the CO). 
Note that in this scenario, the global SED fits (which usually include the supposed UV/optical counterpart) would be called into question.

For GN20.2b, the relatively poorly--constrained redshift of the supposed optical counterpart ($z > 3.2$) makes it difficult to say whether it is related to the now more robustly--detected CO source.
Assuming it is related, its large reddening suggests it is more highly obscured than GN20.2a \citep{2009ApJ...694.1517D}.
For GN20.2a, the spectroscopic redshift for the optical counterpart makes it likely that it is physically associated with the CO emission.
The ACS source to the Northwest of the cross in the GN20.2a map is also a $B$--band Lyman--break galaxy, which \citet{2009ApJ...694.1517D} speculated may be participating in a major merger with the supposed companion LBG.
This could be true whether the companion LBG is a distinct galaxy or an unobscured gap in a large dust screen.
Overall, the presence of more than one Lyman break galaxy within $\sim$1$^{\prime\prime}$ of the CO,
combined with the more compact (than GN20.2b) CO emission, larger gas surface density, larger FWHM linewidth, and smaller implied obscuration \citep{2009ApJ...694.1517D} 
may indicate that GN20.2a is in a different merging stage than GN20.2b.

\begin{figure*}
\centering
\includegraphics[scale=0.6]{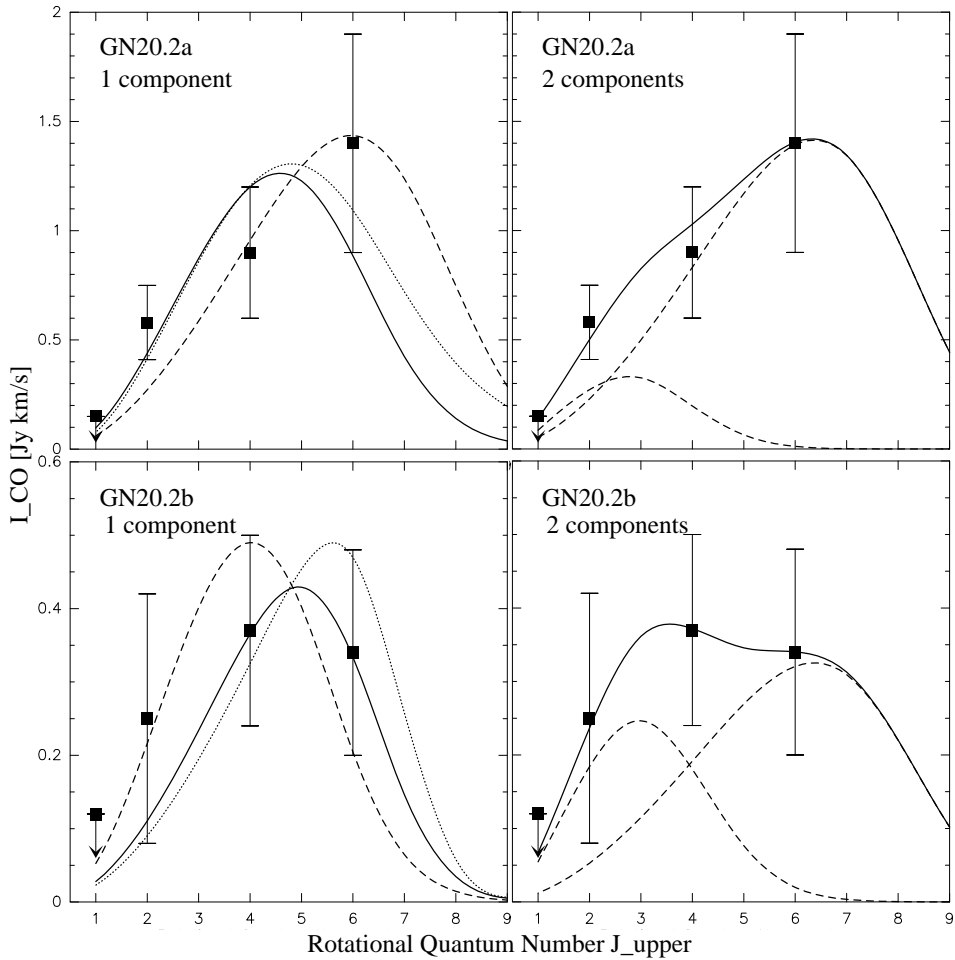}
\caption{CO SLEDs for the SMGs GN20.2a (top) and GN20.2b (bottom). We fit the data for each SMG with a range of one--component LVG models (left) as well as a two--component model (right).
For GN20.2a, the one--component models have kinetic temperatures/H$_2$ densities of 65K/3200 cm$^{-3}$ (solid), 65K/7900 cm$^{-3}$ (dotted), and 200K/2000 cm$^{-3}$ (dashed), while the two--component model has one component with 20K/1000 cm$^{-3}$ and one component with 65K/10,000 cm$^{-3}$.
For GN20.2b, the one--component models shown are 25K/15,800 cm$^{-3}$ (dotted),
35K/6300 cm$^{-3}$ (solid),
and 45K/2500 cm$^{-3}$ (dashed),
and the two component model has one component with 30K/1000 cm$^{-3}$ 
and one component with 65K/10,000 cm$^{-3}$.
See text for further details.
Note that the CO(1--0) values for GN20.2a and GN20.2b are upper limits.
}
\label{fig:sled}
\end{figure*}

\subsubsection{CO Excitation}
\label{sled}

The spectral line energy distributions (SLEDs) for CO in GN20.2a and GN20.2b are shown in Figure~\ref{fig:sled}. 
We modeled these data with standard radiative transfer, large velocity gradient models \citep[LVG;][]{1974ApJ...187L..67S} in order to constrain the physical conditions of the gas in these SMGs.
The collision rates were taken from \citet{0953-4075-34-13-315}, with an ortho--to--para ratio of 3:1 and a fixed CO abundance per velocity gradient of [CO]/$\Delta v$ = 1 $\times$ 10$^{-5}$ pc (km s$^{-1}$)$^{-1}$ \citep[e.g.,][]{2010ApJ...714.1407C, 2011ApJ...739L..31R}.
This analysis yields information on the gas temperature, density, and filling factor, assuming the CO is distributed in a face--on filled circular disk of a fixed size.
As our measured I$_{\rm CO}$ values carry significant uncertainties, and given the known degeneracy between kinetic temperature and density, 
we show a range of single component gas models which can fit the data for each galaxy.
For GN20.2a, fixing the size to that measured from the CO(2--1) (Section \ref{COsize}), the three gas models shown are:
1) a kinetic temperature of 65K, an H$_2$ density of 3200 cm$^{-3}$, and a filling factor of 0.19, 
2) a kinetic temperature of 65K, an H$_2$ density of 7900 cm$^{-3}$, and a filling factor of 0.09, and
3) a kinetic temperature of 200K, an H$_2$ density of 2000 cm$^{-3}$, and a filling factor of 0.15.
Thus, the CO data can be described by a large range of conditions, though none of the fits are particularly good.
We also fit the data with a two--component gas model, as has sometimes been required for SMGs \citep{2010ApJ...714.1407C, 2011ApJ...739L..31R, 2013MNRAS.429.3047B}.
In this case, the lower excitation component has a kinetic temperature of 20K, an H$_2$ density of 1000 cm$^{-3}$, and a filling factor of 1.3.
(We note that a filling factor $>$1 is unphysical, but as we are using the half--light radii to estimate the size, the filling factor is actually $<$1 over the full extent of the disk.)
The higher excitation component for GN20.2a has a kinetic temperature of 65K, an H$_2$ density of 10,000 cm$^{-3}$, and a filling fator of 0.07.
Here, we fixed the size of the higher excitation component to the CO(2--1) size as well since the spatial extents of the higher-excitation lines are relatively unconstrained.
In this model, therefore, the low excitation component would contribute more than half of the CO(1--0) and CO(2--1) emission, decreasing in importance for higher-J lines.

For GN20.2b, the three single--component models shown are:
1) a kinetic temperature of 25K, an H$_2$ density of 15,800 cm$^{-3}$, and a filling factor of 0.04, 
2) a kinetic temperature of 35K, an H$_2$ density of 6300 cm$^{-3}$, and a filling factor of 0.03, and
3) a kinetic temperature of 45K, an H$_2$ density of 2500 cm$^{-3}$, and a filling factor of 0.06.
All models use the average size measured from the CO(2--1) emission.
The range in possible temperatures here is more well constrained, and all models are lower-excitation than the single--component GN20.2a models.
The two--component model for GN20.2b includes a lower excitation component with a kinetic temperature of 30K, an H$_2$ density of 1000 cm$^{-3}$, and a filling factor of 0.15, and a higher excitation component with a kinetic temperature of 65K, an H$_2$ density of 10,000 cm$^{-3}$, and a filling factor of 0.006. 
This would imply that the low--excitation component dominates the emission for transitions all the way up to CO(4--3). 

While it is difficult to put more stringent constraints on the physical conditions of the gas in GN20.2a and GN20.2b, the range of models explored here implies that their excitation properties may be generally consistent with that observed in other SMGs \citep[e.g.,][]{2010ApJ...714.1407C, 2011ApJ...739L..31R, 2013MNRAS.429.3047B} which are typically thought to display moderate excitation
and often require more than one gas component.
In contrast, significantly higher excitation gas has been observed in high--redshift quasar host galaxies and the nuclear starburst regions of nearby galaxies \citep[e.g.,][]{2003ApJ...586..891B, 2004A&A...427...45B, 2006ApJ...650..604R}. 
In all of the possible models, GN20.2b displays lower excitation than GN20.2a, perhaps consistent with the more diffuse CO(2--1) reservoir observed. 
In the two--component model, its low excitation component is even more dominant than the low excitation components found in other SMGs \citep[e.g.,][]{2010ApJ...714.1407C, 2011ApJ...739L..31R},
suggesting its excitation properties 
may even be more comparable to the low excitation seen in the inner disk of the Milky Way and normal star--forming galaxies at $z\sim2$ 
\citep{2009ApJ...698L.178D, 2010ApJ...708L..36A}.  
Further observations of the CO emission in these SMGs, including the CO(5--4) and CO(7--6) lines, will be crucial in confirming these results.

\subsection{A Molecular gas--rich proto--cluster?}
\label{field}

In addition to studying the known CO sources in the field, we have used the deep CO(2--1) VLA data to do both a targeted search 
at the positions of the known $B$--band dropout LBGs 
and a blind search for any emission line sources in the data cube.
Aside from the LBGs associated with GN20 and GN20.2a, and one LBG which is confused with (or possibly merging with -- See Section~\ref{localization}) GN20.2a, 
we detect only one other possible line in the spectrum of 
LBG \#8 (Figure~\ref{fig:bdrop11+HST}).
If this weak ($\sim$4$\sigma$) line is real,
and if it corresponds to CO(2--1), 
then the implied total gas mass is 1.2$\times10^{10}$ ($\alpha_{\rm CO}/0.8$) M$_{\sun}$.
Taking the stellar mass as the average stellar mass of the 77 IRAC--detected $B$--band dropout LBGs in the GOODS-N field \citep[1.5 $\times$ 10$^{10}$ M$_{\odot}$;][]{2009ApJ...694.1517D},
the baryonic gas mass fraction $f_{\rm gas} = M_{\rm gas} / (M_{\rm gas} + M_{\rm stars})$ is $\sim$45\%, 
the same as that reported for the CO(1--0)--detected $z\sim3$ LBG MS 1512-cB58 using the same conversion factor \citep{2010ApJ...724L.153R}.
It is likely that the conversion factor for these galaxies is higher than the ULIRG--like value assumed, as $\alpha_{\rm CO}$ is generally thought to increase for sources with lower gas surface densities \citep{1998ApJ...507..615D, 1997ApJ...484..702S, 2008ApJ...680..246T}.
If we were to instead assume a value of $\alpha_{\rm CO}$ $=$
3.6 M$_{\sun}$ (K km s$^{-1}$ pc$^{2}$)$^{-1}$, as reported for $z=1.5$ star forming disk galaxies \citep[BzKs;][]{2010ApJ...713..686D},
then the gas mass would increase by a factor of $\sim$4.

The derived redshift of this line (4.0452$\pm$0.0004) indicates a radial distance from GN20 of 8 comoving Mpc.
While the transverse size of the GN20 proto--cluster has been suggested to be 2 comoving Mpc, its radial size is basically unconstrained due to errors on the redshift estimates and the possibility of peculiar motions in the radial direction \citep{2009ApJ...694.1517D}.
If the detected line is real, this source may be associated with the proto--cluster.

Nothing was detected for the remaining LBGs.
Aside from one LBG (\#6 in Figure~\ref{fig:bdrops_130kL}; a.k.a.\ BD29079) with a spectroscopic redshift placing it definitively in our bandpass \citep[$z=4.058$;][]{2009ApJ...694.1517D},
the exact redshifts of the rest of the LBGs are unknown,
and their redshifted CO emission may lie outside the range covered by our observations ($z=$4.035--4.063).
For BD29079 and any other sources which \textit{may} be covered by our observations,
our non--detections of CO emission allow us to constrain their luminosities.
If we assume a linewidth of 300 km s$^{-1}$ (FWHM),
then we derive 3$\sigma$ upper limits on their CO luminosities in the range (8--16) $\times$ 10$^{9}$ K km s$^{-1}$ pc$^{2}$ (depending on the primary beam correction at their position).
The stellar mass for BD29079 is $2.6 \times 10^{10}$ M$_{\odot}$, 
and for the rest of the LBGs, we assume the average stellar mass reported for 77 IRAC--detected $B$--band dropout LBGs in the GOODS-N field
\citep[as well as assumed above for the possible emission line source LBG \#8;][]{2009ApJ...694.1517D}.
This implies that, on average, these LBGs have gas--to--stellar mass ratios $<0.4-0.9 \times (\alpha_{\rm CO}/0.8)$
and average gas mass fractions of $<$30--45\% (or $<$65--80\% if $\alpha_{\rm CO} = 3.6$).
For BD29079, the gas--to--stellar mass ratio is $<0.35 \times (\alpha_{\rm CO}/0.8)$, and the average gas mass fraction is $<$26\% (or $<$60\% if $\alpha_{\rm CO} = 3.6$).
We therefore cannot say with any certainty whether the strong evolution in the molecular gas content reported for $z\sim1.5$ galaxies \citep{2010ApJ...713..686D} continues up to $z\sim4$.

Taking this argument in the opposite direction, 
if we assume these LBGs are similar to the $z=1.5$ BzK galaxies of \citet{2010ApJ...713..686D}, 
with gas fractions as large as 65\% 
and a conversion factor of 3.6 M$_{\sun}$ (K km s$^{-1}$ pc$^{2}$)$^{-1}$,
then a handful of them (including BD29079) would have been just below our detection threshold.
However, it has been argued that the conversion factor increases for objects with lower metallicities \citep{2012ApJ...746...69G, 2012MNRAS.421.3127N},
and that the cosmic evolution of the mass--metallicity relation generally favors lower metallicities at higher redshift \citep{2008A&A...488..463M}.
Tan et al. (in prep) quantifies and constrains this effect within the current understanding of metallicity evolution at high--z.
With this in mind, these $z\sim4$ LBGs may well have an even higher conversion factor than the assumed ($z=1.5$ BzK) value.
It is no surprise, then, that we do not detect any strong CO emission from the LBGs.

The blind search also failed to uncover any unambiguous new sources of CO emission.
The deepest search produced 
$\sim$5$\pm$1 emission line source candidates
(within our limited statistics),
two of which correspond to the known sources GN20 and GN20.2a.
Of the four previously unidentified source candidates,
only one has a possible counterpart, which would place its redshift at $z\sim1.5$ (i.e.\ unrelated to the GN20 proto--cluster).
The remaining sources have no counterparts in the deep HST$+$ACS imaging, suggesting that (if they are real) their stellar light is 
entirely obscured.
This fact alone calls their reality into question,
as such significant dust obscuration is typically only seen in bright SMGs.

If they were real and associated with the $z=4.05$ proto--cluster in the field, 
then the detected emission would be CO(2--1) 
and their total gas masses would be $0.9-1.8\times10^{10}$ ($\alpha_{\rm CO}/0.8$) M$_{\sun}$.
Their redshifts would be in the range $z = 4.0377-4.0601$, implying radial distances from GN20 of 2.5--14 comoving Mpc (with an average uncertainty of $<$1 Mpc).
This could imply that they are associated with the proto--cluster.
Whether or not they are real, however, is a question that will have to await further follow-up observations.

\section{CONCLUSIONS}
\label{conclusions}

In conclusion, we presented a study of the molecular gas in the GN20 proto--cluster at $z=4.05$
via spectroscopic imaging of the CO emission.
Using a uniquely deep CO(2--1) dataset, we resolved the gas in member SMGs GN20.2a and GN20.2b on scales down to 1.3 kpc.
We measured a CO(2--1) deconvolved size of $\sim$5 $\times$ 3 kpc for GN20.2a (Gaussian FWHM, projected),
significantly smaller than the very extended ($\sim$15 kpc) reservoirs measured for some SMGs in CO(1--0).
If such extended gas reservoirs are typical of the SMG population as a whole, 
this indicates either GN20.2a is more compact than most, 
or that the low--J CO(2--1) emission is still not tracing the full extent of the gas reservoir.
In GN20.2b, on the other hand, we see a more extended gas reservoir ($\sim$8 $\times$ 5 kpc), with a size roughly double that of typical CO(3--2)/CO(4--3) sizes. 

The average gas surface densities for GN20.2a and GN20.2b are $\sim$3900$\times$(sin \textit{i}) ($\alpha_{\rm CO}$/0.8) M$_{\sun}$ pc$^{-2}$ and $\sim$530$\times$(sin \textit{i}) ($\alpha_{\rm CO}$/0.8) M$_{\sun}$ pc$^{-2}$, respectively, considerably higher than the densities observed in local spiral galaxies and GMCs.
At higher resolution, these values increase to $\sim$12,700$\times$(sin \textit{i}) ($\alpha_{\rm CO}$/0.8) M$_{\sun}$ pc$^{-2}$ and $\sim$1700$\times$(sin \textit{i}) ($\alpha_{\rm CO}$/0.8) M$_{\sun}$ pc$^{-2}$ for the most compact components.
While the average surface density of GN20.2b is comparable to normal star forming galaxies at $z\sim1-2$,
the extremely high values seen in GN20.2a may require a different mechanism such as a major merger.

We used the published SFR for GN20.2a to estimate an average SFR density of $\sim$80$\times$(sin \textit{i}) M$_{\sun}$ yr$^{-1}$ kpc$^{-2}$. This is well below that expected for Eddington--limited maximal starbursts.
Its gas consumption timescale is much shorter than normal star forming galaxies, at 50 ($\alpha_{\rm CO}$/0.8) Myr.

Using the FWHM linewidth of the CO(2--1) emission, we estimated a dynamical mass for GN20.2a of (2.2$\pm$0.9) $\times$ 10$^{11}$ M$_{\sun}$ within $R \sim 2.5$ kpc.
Assuming 25\% dark matter and equal gas and stellar masses, 
we derived a CO--to--H$_2$ conversion factor of $\alpha_{\rm CO}$ $=$ 1.7$\pm0.8$ M$_{\sun}$ (K km s$^{-1}$ pc$^{2}$)$^{-1}$.
GN20.2b is less well--constrained, but its estimated dynamical mass of (6$\pm^{7}_{6}$) $\times$ 10$^{10}$ M$_{\sun}$ within $R \sim 4$ kpc
implies a CO--to--H$_2$ conversion factor of $\alpha_{\rm CO}$ $=$ 1.1$\pm^{1.5}_{1.1}$ M$_{\sun}$ (K km s$^{-1}$ pc$^{2}$)$^{-1}$.
Both values may therefore support the assumption of a low, ULIRG--like conversion factor for these systems, agreeing with estimates for the nearby SMG (and fellow proto--cluster member) GN20 by \citet{2012ApJ...760...11H} and \citet{2011ApJ...740L..15M}, although many assumptions went into these estimates. 

We found evidence for significant offsets (0.5$^{\prime\prime}$--1$^{\prime\prime}$) between the CO emission of GN20.2a and GN20.2b and their presumed HST$+$ACS counterparts.
This may indicate the presence of a large dust screen coincident with the CO emission and blocking the majority of the UV/optical light.
Alternately, the optical counterparts may be distinct galaxies from the dusty starburst galaxies emitting in CO.
In the case of GN20.2a, the presence of a second nearby HST$+$ACS source, along with its compact size, higher gas surface density, very broad linewidth, and lower implied obscuration may indicate that it is in a different merging stage than GN20.2b.

By combining our CO(2--1) data with VLA and PdBI datasets targeting other CO transitions, we constructed CO spectral line energy distributions for GN20.2a/b. 
Fitting the data with a range of one-- and two--component LVG models, we found that their excitation properties may be generally consistent with that observed in other SMGs.
GN20.2b displays lower excitation than GN20.2a, again consistent with the conclusion that the two SMGs are in different merging stages.

In addition to studying the known SMGs in the field,
we carried out a targeted search for CO emission at the positions of 14 $B$--band dropouts, tentatively detecting an emission line in a previously--undetected LBG. 
This emission line would imply a source redshift of 4.0452$\pm$0.0004, assuming it is CO(2--1), and a total gas mass of 1.2 $\times$ 10$^{10}$ ($\alpha_{\rm CO}/0.8$) M$_{\odot}$.
No emission was detected from the remaining LBGs, though the lack of spectroscopic redshifts for all but one source (BD29079) mean that we cannot be sure that they fell within our bandpass.
For BD29079 and any other sources in the correct redshift range, we placed 3$\sigma$ upper limits on their CO luminosities of (8--16) $\times$ 10$^{9}$ K km s$^{-1}$ pc$^{2}$.
Even if they have gas fractions as high as $z\sim1.5$ BzK galaxies, their CO--to--H$_2$ conversion factor is likely higher, meaning that we would not expect to detect them with these limits.

Finally, we carried out a blind search for emission--line sources down to a 5$\sigma$ limiting CO luminosity of $L^{\prime}_{\rm CO(2-1)} = 8 \times 10^{9}$ K km s$^{-1}$ pc$^{2}$ and covering $\Delta z = 0.0273$ ($\sim$20 comoving Mpc) at $z\sim$4.05.
The search produced $\sim$5$\pm$1 emission line candidates, two of which are known sources. 
If the emission lines are real and correspond to CO(2--1), the sources have redshifts in the range $4.0377-4.0601$ and total gas masses of $0.9-1.8$ $\times$ 10$^{10}$ ($\alpha_{\rm CO}/0.8$) M$_{\odot}$.
Only one of the remaining source candidates has an optical counterpart, and its photometric redshift ($z = 1.5$) implies that it is unrelated to the proto--cluster.
Therefore, we did not detect any other strong, unambiguous sources of CO emission associated with the $z\sim4$ proto--cluster.

\acknowledgements
The authors wish to thank Roberto Decarli, Qinghua Tan, and Lindley Lentati for assistance with various aspects of the data reduction/analysis, 
and we thank Christian Henkel for his work on the LVG code.
The authors also wish to thank the anonymous referee for helpful comments which led to the improvement of this paper.
CC thanks the Kavli institute of Cosmology for their hospitality.
ED acknowledges support from grants ERC--StG UPGAL 240039 and ANR-08-JCJC-0008.
DR acknowledges funding from NASA through a Spitzer Space Telescope grant. 
The National Radio Astronomy Observatory is a facility of the National Science Foundation under cooperative agreement by Associated Universities, Inc.


\bibliographystyle{apj}
\bibliography{Hodge}

\end{document}